\documentclass[12pt]{article}
\usepackage{amsmath}
\usepackage{amssymb}
\usepackage{amsfonts}
\usepackage{graphicx}
\usepackage{natbib}  
\usepackage{geometry}
\usepackage[hyphens]{url}
\usepackage{microtype}

\title{Statistical Comparison of Hidden Markov Models via Fragment Analysis}

\author{Carlos M. Hernandez-Suarez\thanks{Coordinaci\'on General de Investigaci\'on Cient\'ifica, Universidad de Colima, Colima, M\'exico. \texttt{cmh1@cornell.edu}} 
\and 
Osval A. Montesinos-L\'opez\thanks{Facultad de Telem\'atica, Universidad de Colima, Colima, M\'exico. \texttt{oamontes1@ucol.mx}}}

\date{\today}
\begin{document}

\maketitle

\newcommand{\diag}{\mathrm{Diag}}

\begin{abstract}
Standard practice in Hidden Markov Model (HMM) selection favors the candidate with the highest full-sequence likelihood, although this is equivalent to making a decision based on a single realization.
We introduce a \emph{fragment-based} framework that redefines model selection as a formal statistical comparison.  
For an unknown true model $\mathrm{HMM}_0$ and a candidate $\mathrm{HMM}_j$, let $\mu_j(r)$ denote the probability that $\mathrm{HMM}_j$ and $\mathrm{HMM}_0$ generate the same sequence of length~$r$.  
We show that if $\mathrm{HMM}_i$ is closer to $\mathrm{HMM}_0$ than $\mathrm{HMM}_j$, there exists a threshold $r^{*}$---often small---such that $\mu_i(r)>\mu_j(r)$ for all $r\geq r^{*}$.  
Sampling $k$ independent fragments yields unbiased estimators $\hat{\mu}_j(r)$ whose differences are asymptotically normal, enabling a straightforward $Z$-test for the hypothesis $H_0\!:\,\mu_i(r)=\mu_j(r)$.  
By evaluating only short subsequences, the procedure circumvents full-sequence likelihood computation and provides valid $p$-values for model comparison.
\end{abstract}


\maketitle
\section{Introduction}

Hidden Markov Models (HMMs) are a cornerstone of time-series modeling, underpinning applications from speech recognition~\citep{rabiner1989tutorial} to bioinformatics and finance. However, selecting an appropriate model among several competing HMMs becomes difficult when the observed sequence is long. Standard criteria such as the Akaike Information Criterion (AIC)~\citep{akaike1974new} or the Bayesian Information Criterion (BIC)~\citep{schwarz1978estimating} rely on exact likelihood calculations of large data sequences, which are both numerically unstable and computationally expensive.

Several proposals seek to mitigate these numerical issues, for instance by normalizing probability vectors or employing logarithmic transformations within the forward-backward algorithm~\citep{rabiner1989tutorial}. Yet, full-sequence likelihoods can become \emph{prohibitively expensive} for tens of thousands (or more) observations. Approximate Bayesian Computation~\citep{beaumont2002approximate,marin2012abc} and spectral approaches~\citep{hsu2012statistical} circumvent explicit likelihood computation by leveraging summary statistics or matrix factorizations. Stochastic Variational Inference~\citep{hoffman2013stochastic} further shows promise for parameter estimation on massive sequences, but direct comparison across candidate HMMs remains problematic. 

To tackle these challenges, \citet{hernandez2024optimizing} introduce a fragment-based or `likelihood-free' framework to compare HMMs without computing their full-sequence likelihoods. The primary idea is: (i) randomly sample short subsequences (fragments) of length $r$ from the data; (ii) evaluate their likelihood under each candidate HMM; and (iii) use standard statistical tools to compare these likelihoods. This approach drastically reduces the computational overhead of typical forward-backward methods over very large sequences. Moreover, it naturally accommodates different numbers of hidden states, which can invalidate standard likelihood ratio tests~\citep{giudici2000} that require nested models.

Here, we formally present the mathematical foundation of the fragment-based metric described in \citep{hernandez2024optimizing}. We define $\mu_j(r)$ as the \emph{expected likelihood} under model~$j$ for a randomly selected fragment of length $r$ generated by the unknown true model, $\mathrm{HMM}_0$. Under suitable conditions, we show that $\mu_j(r)$ can serve as a reliable surrogate for the full-sequence likelihood in large-scale HMM comparisons. Additionally, we describe a straightforward procedure for constructing candidate HMMs that replicate the \emph{observed-state transition frequency matrix} derived from data, thereby ensuring consistency in the short-fragment regime.

\section{Notation and Preliminaries}
\label{sec:notation}

We consider a family of Hidden Markov Models (HMMs) each with $N$ hidden states and $K$ observable states (or symbols). For a generic model $\mathrm{HMM}_j$, let
\begin{itemize}
\item $\mathbf{P}_j \in \mathbb{R}^{N\times N}$ denote its hidden-state transition matrix (row-stochastic),
\item $\mathbf{S}_j \in \mathbb{R}^{N\times K}$ denote its emission matrix (rows summing to 1),
\item $\boldsymbol{\pi}_j \in \mathbb{R}^N$ denote the stationary distribution of $\mathbf{P}_j$, satisfying $\boldsymbol{\pi}_j^\top \mathbf{P}_j = \boldsymbol{\pi}_j^\top$.
\end{itemize}

Let $X_n = (x_1, \dots, x_n)$ denote the hidden-state chain and $Y_n = (y_1, \dots, y_n)$ the observed sequence, each $y_t \in \{1,\dots,K\}$. We assume the data $Y_n$ arise from an unknown \emph{true} model $\mathrm{HMM}_0$ with associated matrices $\mathbf{P}_0$, $\mathbf{S}_0$, and $\boldsymbol{\pi}_0$. 

We define a \emph{fragment} of length $r$ as any contiguous subsequence $(y_\ell,\dots,y_{\ell +r-1})$ of $Y_n$. Since $K^r$ is the total number of distinct fragments of length $r$, we may index such possible fragments by $s_i$, $i=1,\dots,K^r$, and denote by $l_j(s_i)$ the likelihood of $s_i$ under $\mathrm{HMM}_j$. 

\section{A New Fragment-Based Metric}
\label{sec:metric}

\subsection{Definition of $\mu_j(r)$}

We are interested in comparing models by their expected likelihood to generate a random fragment of length $r$. When $Y_n$ is truly generated by $\mathrm{HMM}_0$, each fragment $s_i$ of length $r$ has probability $l_0(s_i)$. Thus the \emph{expected fragment likelihood} under $\mathrm{HMM}_j$ is
\begin{equation}
\label{eq:muJR}
\mu_j(r) = \sum_{i=1}^{K^r} l_0(s_i)\, l_j(s_i)
\end{equation}
which is also the probability that two length-$r$ sequences, each independently generated by $\mathrm{HMM}_0$ and $\mathrm{HMM}_j$, respectively, coincide exactly. Because $l_0(s_i)$ is the pmf for the $s_i$-indexed random fragment, we have $\mu_j(r) = \text{E}[L_j(r)]$, where $L_j(r)$ is the likelihood of a randomly drawn fragment of size $r$.

An equivalent matrix form uses the so-called \emph{forward} factorization:
\begin{equation}
\label{eq:muX}
l_j(s_i) 
= 
\boldsymbol{\pi}_j^\top\,\mathbf{M}_j(y_1)\,\mathbf{M}_j(y_2)\,\cdots\,\mathbf{M}_j(y_r)\,\mathbf{1},
\end{equation}
with 
\[
\mathbf{M}_j(m) \;=\; \mathbf{P}_j\,\diag\!\bigl(\mathbf{S}_j(\cdot,m)\bigr).
\]
Summing over all $K^r$ fragments, the Kronecker product gives a compact form for $\mu_j(r)$. Specifically,
\begin{eqnarray}
\label{kron}
\mu_j(r) 
&=& 
\left( \boldsymbol{\pi}_0 \otimes \boldsymbol{\pi}_j \right) 
\biggl( \sum_{i=1}^{K} \mathbf{M}_0(i) \otimes \mathbf{M}_j(i) \biggr)^r 
\mathbf{1}
\nonumber 
\\
& = & 
\left( \boldsymbol{\pi}_0 \otimes \boldsymbol{\pi}_j \right) 
\mathbf{W}_j^r\, \mathbf{1},
\end{eqnarray}
where 
\begin{equation}
\label{eq:Wr}
\mathbf{W}_j = 
\left( \mathbf{P}_0 \otimes \mathbf{P}_j \right) 
\cdot 
\operatorname{Diag} \left( \operatorname{vec} \left( \mathbf{S}_j \mathbf{S}_0^\top \right) \right)
\end{equation}

\section{Testing Based on Sampled Fragments}

Given two candidates $\mathrm{HMM}_1$ and $\mathrm{HMM}_2$, we aim to compare 
\[
\mu_1(r)\quad \text{vs.} \quad \mu_2(r).
\] 
Generally, we do not know $\mathrm{HMM}_0$, so we draw $k$ fragments $\{s_1,\dots,s_k\}$ of length $r$ from the observed sequence $Y_n$. For each $s_i$, compute $l_1(s_i)$ and $l_2(s_i)$, then set $d_i = l_1(s_i) - l_2(s_i)$. By the Central Limit Theorem (CLT), under mild conditions,
\begin{equation}
\label{eq:dif_mean}
\bar{d} \;=\; \frac{1}{k}\,\sum_{i=1}^k\,d_i
\;\;\;\;\xrightarrow[]{d}\;\;\;
N\!\Bigl(\mu_1(r) - \mu_2(r),\;\tfrac{1}{k}\,\sigma_{12}^2(r)\Bigr),
\end{equation}
where 
\begin{equation}
\label{eq:variance}
\sigma_{12}^2(r)  
= 
\mu_{12}^2(r) \;-\; \bigl(\mu_{12}(r)\bigr)^2,
\quad
\text{with}
\quad
\mu_{12}(r)=\mu_1(r)\;-\;\mu_2(r).
\end{equation}

\noindent An explicit form for $\mu_{12}^2(r)$ is given by:
\begin{align*}
\mu_{12}^2(r)\;=&\;\left( \boldsymbol{\pi}_0 \otimes \boldsymbol{\pi}_1 \otimes \boldsymbol{\pi}_1 \right)
\left( \sum_{i=1}^{K} \mathbf{M}_0(i) \otimes \mathbf{M}_1(i) \otimes \mathbf{M}_1(i) \right)^r \mathbf{1} 
\\
&\;-\;2\;\left( \boldsymbol{\pi}_0 \otimes \boldsymbol{\pi}_1 \otimes \boldsymbol{\pi}_2 \right)
\left( \sum_{i=1}^{K} \mathbf{M}_0(i) \otimes \mathbf{M}_1(i) \otimes \mathbf{M}_2(i) \right)^r \mathbf{1} 
\\
&\;+\;\left( \boldsymbol{\pi}_0 \otimes \boldsymbol{\pi}_2 \otimes \boldsymbol{\pi}_2 \right)
\left( \sum_{i=1}^{K} \mathbf{M}_0(i) \otimes \mathbf{M}_2(i) \otimes \mathbf{M}_2(i) \right)^r \mathbf{1}.
\end{align*}
Hence, we can test 
\begin{equation}
\label{eq:H0}
H_0: \mu_1(r) = \mu_2(r)\quad \text{vs.}\quad H_1: \mu_1(r) > \mu_2(r)
\end{equation}
by a standard $Z$-statistic:
\begin{equation}
\label{eq:Zk}
Z(k)\;=\;\frac{\bar{d}}{\hat{S}_d/\sqrt{k}},
\end{equation}
where $\hat{S}_d^2$ is the sample variance of the $\{d_i\}$. 

\subsection{On the Optimal Fragment Size r}
 In the long run, as $r\to\infty$, each $\mu_j(r)$ grows or decays according to the spectral radius of its associated matrix $\mathbf{W}_j$~\eqref{eq:Wr}. Thus, while $r$ can be increased until dominance becomes evident, a practical strategy is to use moderate $r$ (e.g., $3\le r \le 6$) in tandem with a sampling-based $Z$-test. Because sampling and fragment likelihood calculations for small $r$ are efficient, one can attempt multiple $r$ values until the comparison stabilizes.

\section{Example}
\label{sec:examples}

The dataset, compiled by the Central Pollution Control Board and publicly available through Kaggle \citep{jha2023ozone}, contains 4,560 consecutive daily observations of atmospheric pollutant concentrations and meteorological variables.  
Only ozone ($\mathrm{O}_3$) concentrations were analyzed to construct and compare candidate HMMs for the underlying air quality dynamics.  
Observations were discretized into three categories---low, medium, and high---based on empirical terciles of the ozone distribution.  
Two candidate Hidden Markov Models (HMMs) were fitted to the categorized daily ozone data using the \texttt{hmmlearn} Python package.  
The Expectation-Maximization (EM) algorithm was employed to maximize the likelihood, allowing up to 1000 iterations and fixing the random seed to ensure reproducibility.  
HMM$_1$ assumed three hidden states and HMM$_2$ assumed four hidden states, both with three observable categories.  
The resulting transition and emission probability matrices were extracted for further analysis and are presented in the Appendix.

Table \ref{tab:fragment_comparison} shows the statistical comparison of both models for fragment sizes $r$ from 3 to 20.  
The sample size in all cases was $n = 1,000$.  The log-likelihood of the entire observed sequence under each model was: HMM$_1$: $-3581.6988$, and HMM$_2$: $-3009.0871$.

\begin{table}
\centering
\begin{minipage}{\textwidth}
\caption{Comparison of HMM$_1$ and HMM$_2$ based on fragment sampling ($n=1,000$ fragments per $r$).}
\label{tab:fragment_comparison}
\begin{tabular*}{\textwidth}{@{}l@{\extracolsep{\fill}}cccc@{}}
\hline
$r$ & $\hat{\mu}_2(r)-\hat{\mu}_1(r)$ & Std Difference & $Z$-statistic & $p$-value \\
\hline
3 & 0.03429         & 0.04805        & 22.567    & $<10^{-7}$ \\
4 & 0.03125         & 0.04215        & 23.444    & $<10^{-7}$ \\
5 & 0.02705         & 0.03621        & 23.619    & $<10^{-7}$ \\
6 & 0.02434         & 0.03225        & 23.868    & $<10^{-7}$ \\
7 & 0.01975         & 0.02864        & 21.809    & $<10^{-7}$ \\
\hline
\end{tabular*}
\end{minipage}
\vspace*{-6pt}
\end{table}

\begin{figure}
 \centerline{\includegraphics[width=6in]{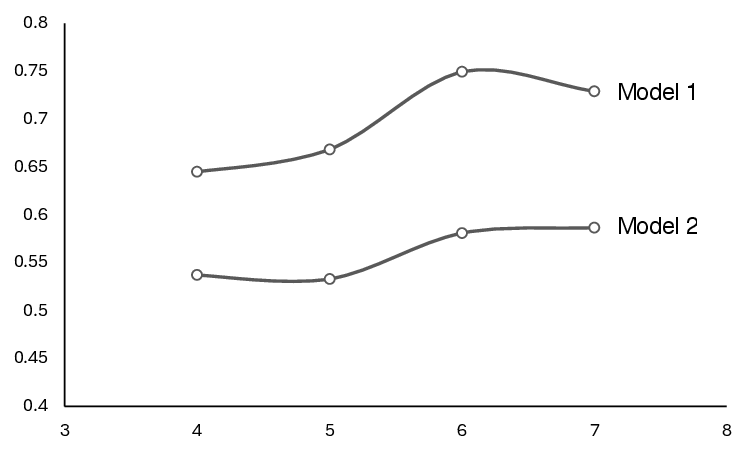}}
\caption{Plot of the data in Table \ref{tab:rates}. No overlap observed in the studied interval.}
\label{f:figure1}
\end{figure}

Table \ref{tab:rates} shows the estimates of the ratios $\mu_i(r+1)/\mu_i(r)$, $i=1,2$, in an effort to illustrate the role of the dominant eigenvalue in governing the asymptotic growth of fragment probabilities as the fragment length $r$ increases. These data is plotted in Figure \ref{f:figure1}.
\begin{table}
\centering
\begin{minipage}{\textwidth}
\caption{Estimates of the dominant eigenvalues, expressed as the ratio $\hat{\mu}_i(r+1)/\hat{\mu}_i(r)$, and the number of possible sequences of length $r$. For both models, the number of observable states is $K=3$. Sequence size is $N =4,560$}
\label{tab:rates}
\begin{tabular*}{\textwidth}{@{}l@{\extracolsep{\fill}}ccccc@{}}
\hline
$r$ & $\hat{\mu}_1(r+1)/\hat{\mu}_1(r)$  & $\hat{\mu}_2(r+1)/\hat{\mu}_2(r)$ & $K^r$ & $K^r/N$\\
\hline
3 &        &        & 27    & 0.0059 \\
4 & 0.5372 & 0.6450 & 81    & 0.0178 \\
5 & 0.5330 & 0.6684 & 243   & 0.0533 \\
6 & 0.5810 & 0.7492 & 729   & 0.1599 \\
7 & 0.5868 & 0.7292 & 2187  & 0.4796 \\
\hline
\end{tabular*}
\end{minipage}
\vspace*{-6pt}
\end{table}

\section{Discussion}
\label{sec:discussion}

Table~\ref{tab:fragment_comparison} shows that at low $r$-values, the null hypothesis $H_0\!:\, \mu_1(r) = \mu_2(r)$ is already rejected, favoring the conclusion that $\mu_1(r) > \mu_2(r)$ for $r \geq 3$. 
While it is theoretically possible for the inequality to reverse at larger $r$-values, depending on the behavior of the dominant eigenvalues of $\mathbf{W}_j$, Table~\ref{tab:rates} and Figure~\ref{f:figure1} indicate that no overlap occurs between the estimated rates $\hat{\mu}_1(r+1)/\hat{\mu}_1(r)$ and $\hat{\mu}_2(r+1)/\hat{\mu}_2(r)$. 
This suggests that a switch in dominance is unlikely. 
Moreover, increasing $r$ beyond a certain point is not advisable, as the sequence length must be much larger than the number of possible fragments of size $r$—that is, $n \gg K^r$—to ensure a representative sampling of the fragment distribution. 
Table~\ref{tab:rates} indicates that this condition is no longer satisfied for $r \geq 6$.

Traditional HMM model selection methods typically rely on the log-likelihood of the entire sequence, which can be numerically unstable or even infeasible for large $n$.  
Although partial normalization steps in the forward-backward algorithm mitigate underflow to some extent, they do not guarantee stable floating-point operations at extreme sequence lengths.  
In contrast, the fragment-based approach evaluates shorter subsequences, keeping computations localized, stable, and computationally efficient.

Moreover, classical likelihood ratio tests~\citep{giudici2000} require nested models and become invalid when candidate HMMs differ in their number of hidden states.  
The fragment-based method overcomes this limitation by allowing direct statistical comparison between arbitrary models, thereby offering greater flexibility for large-scale or exploratory analyses.

A key practical question concerns the choice of fragment size $r$.  
While small $r$ values avoid data sparsity across the $K^r$ possible fragment types, larger $r$ may be necessary to detect more subtle differences between models.  
A straightforward strategy is to run the sampling-based test sequentially at $r=3,4,5,\dots$ until the resulting $Z$-statistics converge to a stable pattern.  
More sophisticated approaches could adaptively tune $r$ based on preliminary significance tests, or incorporate blockwise sampling to account for correlations between neighboring fragments.

Although it is possible for an HMM with \emph{mismatched} emission probabilities to achieve a high $\mu_j(r)$ for some small $r$, the method's reliance on multiple fragment sizes and on the properties of the true data-generating process $\mathrm{HMM}_0$ typically penalizes such mismatches as $r$ increases.  
Future work could investigate the rate at which $\mu_j(r)$ distinguishes the true model as $r$ grows, or explore alternative short-fragment metrics—such as approximate Kullback–Leibler divergences—beyond raw likelihood comparisons.

\section{Conclusion}

We have presented a systematic, fragment-based approach for comparing Hidden Markov Models without requiring full-sequence likelihood computations.  When the emission matrix is unbounded, for instance,  $\mathbf{S}_j$ is a Poisson distribution with parameter $\lambda_j$ all the results still apply because~\eqref{eq:dif_mean} still holds. 

By focusing on short subsequences, the framework is robust to common numerical instabilities in long products of probabilities, and it accommodates models with distinct hidden-state dimensions.  
Importantly, the method enables formal statistical comparisons between models by providing $p$-values that quantify the strength of evidence favoring one model over another.  
This adds a rigorous measure of certainty to model selection, addressing a major limitation of traditional approaches that relied solely on likelihood values without an assessment of statistical significance.  
The method relies on standard sampling and a straightforward test statistic, making it easy to implement on large data sets.  
Future work includes data-driven selection of the fragment length, more refined asymptotic analyses, and further applications in domains where the HMM dimension is large or the observational alphabet is high-dimensional.
\appendix

\section{Fitted HMM Transition and Emission Matrices}

The estimated transition and emission probability matrices for the two candidate HMMs are presented below.

\subsection{Model 1: 3 Hidden States, 3 Observable Categories}

\textbf{Transition Matrix (P1):}
\[
\mathbf{P}_1 = 
\begin{pmatrix}
0.001532 & 0.998364 & 0.000104 \\
0.915998 & 0.014236 & 0.069766 \\
0.027333 & 0.014714 & 0.957953 \\
\end{pmatrix}
\]

\textbf{Emission Matrix (S1):}
\[
\mathbf{S}_1 =
\begin{pmatrix}
0.611939 & 0.383932 & 0.004129 \\
0.597684 & 0.390171 & 0.012145 \\
0.009617 & 0.269177 & 0.721207 \\
\end{pmatrix}
\]

\subsection{Model 2: 4 Hidden States, 3 Observable Categories}

\textbf{Transition Matrix (P2):}
\[
\mathbf{P}_2 = 
\begin{pmatrix}
0.927025 & 0.005267 & 0.067344 & 0.000363 \\
0.005510 & 0.921437 & 0.068471 & 0.004582 \\
0.066956 & 0.069051 & 0.863918 & 0.000075 \\
0.145968 & 0.120110 & 0.731595 & 0.002327 \\
\end{pmatrix}
\]

\textbf{Emission Matrix (S2):}
\[
\mathbf{S}_2 =
\begin{pmatrix}
0.004554 & 0.091130 & 0.904317 \\
0.914271 & 0.085396 & 0.000332 \\
0.102241 & 0.811204 & 0.086555 \\
0.353744 & 0.646253 & 0.000003 \\
\end{pmatrix}
\]

\end{document}